\documentclass[prd, amsfonts, twocolumn, nofotinbib, showpacs, superscriptaddress]{revtex4}
\usepackage{enumerate}
\usepackage{graphicx, epsfig}
\begin{document}
\title{A Dynamical model for non-geometric quantum black holes }

\author{Euro Spallucci}
\thanks{spallucci@ts.infn.it} 
\affiliation{Dipartimento di Fisica, Universit\`a di Trieste and INFN, Sezione di Trieste, Trieste, Italy}

\author{Anais Smailagic}
\thanks{anais@ts.infn.it} 
\affiliation{Dipartimento di Fisica, Universit\`a di Trieste and INFN, Sezione di Trieste, Trieste, Italy}

\begin{abstract} 
It has been recently proposed that quantum black holes can be described as N-graviton Bose-Einstein 
condensates. In this picture the quantum properties of BHs ``... can be understood in terms of the single 
number $N$''\cite{Dvali:2011aa}. However, so far, the dynamical origin of the occupational number $N$ has not been specified.
This description is alternative to the usual one, where black holes are believed to be well described geometrically
 even at the quantum level. 
In  this paper we pursue the former point of view and develop   a non-geometrical, dynamical, model of quantum 
black holes (BHs). In our model the occupational number $N$ turns out to be the principal quantum number $n$
of a Planckian harmonic oscillator. The so-called ``\emph{classicalization}'' corresponds to the large-$n$ limit,
where the Schwarzschild horizon  is recovered.
\end{abstract}

\pacs{04.60.Bc, 13.85.Lg}
\maketitle

	\section{Introduction}
	
         Quantum black holes (BHs) have become an important subject of research since the original paper by Hawking, where the 
         decay, through thermal radiation of nuclear size BHs was shown for the first time.  Nevertheless, in spite of their 
         sub-atomic size the widely accepted view is that they remain classical objects described by solutions of the Einstein 
         equations. Only  matter is treated in terms of quantum field theory. This is known as the ``semi-classical'' approach 
         and draws its justification from the  common believe  that
         quantum gravity itself is relevant only at the Planck scale. The outcome of this approach is that no plausible 
         description of the final BH phase is available in this scenario, leading to an endless, and unsettled, debate on
         the ``information paradox'' \cite{Dvali:2015aja,Hawking:2015qqa,Calmet:2014uaa,Mathur:2013fpa}.\\
         In this framework the most promising approach to the resolution of all above problems seemed to be offered by string theory,
         which is, so far, the only self-consistent, anomaly-free, theory incorporating gravity at the quantum level. 
         Nevertheless, in spite of the successful counting of BH quantum degrees
	 of freedom, consistent with area law,  the actual calculations still rely on  
         stringy improved \emph{classical metrics} \cite{book}. 
         
A different point of view is based on the idea that QBHs are not describable in terms of classical geometry \cite{Dvali:2011aa}.
A completely new, genuinely quantum, approach is needed which only in, a properly defined classical limit, will reproduce the 
known geometrical properties. The above point of view has been laid down in 
\cite{Dvali:2011aa,Dvali:2012mx,Dvali:2012en,Casadio:2013hja,Casadio:2013iqc,Dvali:2014ila}, where a quantum BH  is a self-sustaining  
bound-state of $N$ soft gravitons. This bound state is described as a potential well of width $\sqrt{N}L_P$ and depth
$\hbar/\sqrt{N}L_P $. The BH total mass energy is given by $M=\sqrt{N} M_P$ and its Hawking temperature is $\hbar/\sqrt{N}L_P$.\\
For the sake of chronological record, very similar, bordering equivalent ideas, have been already presented,
many years ago in \cite{Spallucci:1977wc}, where small BHs  were described as  in a purely quantum mechanical framework 
in terms of a potential wells of finite depth, and width corresponding to the classical horizon radius.\\
In this paper, we dig out and update the same idea in view of the recent developments in quantum gravity based on the
celebrated \emph{Holographic Principle} \cite{Susskind:1994vu,Susskind:1998dq,Hooft:1999bw}.\\
It affirms that the dynamics of a general quantum gravitational system \emph{is all encoded into its boundary}.
One of the most transparent realization of this principle is through the the famous Area Law relating the entropy of
a semi-classical BH to  its horizon area.\\
In this paper, we shall promote a classical horizon to a genuine quantum degree of freedom and 
construct its wave equation. In other words, a classically non-dynamical, well define, boundary is allowed to undergo
 quantum fluctuations.  One may expect that a genuine quantum mechanical approach will give raise to a suitable quantum
number that will be finally identified with the occupational number $N$ in \cite{Dvali:2011aa}. \\
            In Section (\ref{pho}) we formulate a classical model for a microscopic, ``breathing'' BH, and solve the equation
            of motion.  This model serves as a starting point for the successive formulation of the quantum wave equation.\\
            In Section (\ref{qh}) we analyze the quantum behavior of the BH horizon and obtain the quantized energy levels.
            Finally, we describe the classical limit of the quantum model.
 
           \section{The ``breathing'' horizon}
         \label{pho}
             
            One of the expectations in future LHC experiments is the appearance of signals indicating the presence
            of Planck scale micro BHs, at least, as virtual intermediate states 
            \cite{Calmet:2008dg,Calmet:2012cn,Calmet:2012fv,Belyaev:2014ljc,Calmet:2014uaa,Calmet:2015fua}. 
            In other words
            they will be just another  structure in the elementary particle zoo. From this point of view, it is hardly
            justifiable that these quantum gravitational excitations can somehow defy the law of quantum mechanics and
            be described on the same geometrical terms as their cosmic cousins of a million, or so, solar masses.
            Oddly enough, so far this has been the dominant point of view. Occasionally, in the distant past a few
            dissonant ideas have been put forward and largely ignored \cite{Markov:1967lha,Markov:1972sc}. 
            Nevertheless, very recently the same line of
            thinking of non-geometric and purely quantum mechanical description  has been proposed as the alternative to 
            the standard view. In order to be completely clear, one is not thinking in terms of any quantum version of Einstein
            General Relativity, but rather in purely particle like quantum mechanical formulation. The classical geometrical
            picture is expected to emerge in a suitable classical limit of the quantum picture. \\
            If the reader really believes that a sub-nuclear size BH can be trustfully described  as a classical solution
            of the Einstein equations, he/she is suggested to stop reading this paper to avoid confusion between geometric and
            non-geometric description of BHs. In what follows
            we shall introduce a particle-like model of micro BHs, where the only link with the classical geometric description is the
            linear relation between its size and total mass-energy.

            \subsection{Classical model}

             In the absence of any quantum equation to start with, one has to develop a suitable quantization procedure of
             some classical, particle-like, model. What should such a classical model represent?\\
              Certainly, not the dynamics of a classical BH seen as a single particle subject to external forces.  
              The main obstacle for quantizing the internal dynamics of a classical BH is that its horizon is a geometrical 
              statical surface. Thus, canonical quantization has no classical counterpart to start with. \\
              Therefore, as a first step, the static horizon has to be given a proper dynamics. In other words, it will
              be assigned its own kinetic energy and allowed to evolve in time.\\
               Furthermore, in the
              case of spherically symmetric BH the problem reduces to the single, radial coordinate which now is allowed
              to ``~\emph{breath}~'' achieving maximum ``~lung capacity~'' at the classical  Schwarzschild 
              radius $r_+= 2M G_N$. \\
               It should be clear by now, that in oder to be able to describe microscopic quantum BHs, which are completely 
              different from cosmic objects, one is forced to leave the safe ground of General Relativity and head towards an
               ``uncharted territory'' . \\
               
               We propose the following relativistic Hamiltonian describing the horizon breathing mode

\begin{equation}
 H\left(\, r\ , p\,\right)\equiv \sqrt{p^2 +\frac{r^2}{4G_N^2} }
\end{equation}

where, $p$ is the canonical radial momentum. As the system is conservative for any classical solution the Hamiltonian is a 
constant of motion:

\begin{equation}
 H=E \label{etot}
\end{equation}

Using the Hamilton equation

\begin{equation}
 \frac{\partial H}{\partial r}=-\dot{p}=\frac{r}{4G_N^2 E}
\end{equation}

together with 

\begin{equation}
 p=\sqrt{E^2 - \frac{r^2}{4G_N^2}} 
\end{equation}
 one finds



\begin{equation}
 \dot{r}^2 =  1 - \frac{r^2}{4G_N^2E^2}
\end{equation}

Choosing the initial condition as:

\begin{equation}
 r\left(\, t=0\,\right)= 0\ .
\end{equation}

one finds the solution:

\begin{eqnarray}
&& r\left(\, t\, \right)= 2G_N E \, \sin\left(\, \omega t\,\right) \\ 
&& \omega\equiv \frac{2\pi}{T}= \frac{1}{2G_N E}
\end{eqnarray}
The oscillation starts from the origin $r =0$ at $t=0$ and reaches the maximum elongation, at the Schwarschild radius
$r_+=2G_N E$, after half a period $t=T/2$. This is possible because we allow a continuous exchange between kinetic and potential
energy, contrary to a static, fixed size, solution of the classical Einstein equations.
 \\
In order to be able to confront  classical and quantum results, to be obtained in the next Section, we calculate the classical 
mean values for $r_+$ and $r_+^2$ defined as time averages over half period  $T/2= \pi G_N E$

\begin{equation}
\widehat{r}\equiv \frac{1}{\pi G_N E}\int_0^{\pi G_N E} dt r(t) =  \frac{4}{\pi} G_N E   =\frac{2}{\pi} r_+ \label{rcl}
\end{equation}

\begin{equation}
 \widehat{r^2}\equiv \frac{1}{\pi G_N E}\int_0^{\pi G_N E} dt r^2(t)=2G_N^2 E^2 = \frac{1}{2}r_+^2 \label{r2cl}
\end{equation}

We see that $\widehat{r} > r_+/2$ contrary to what one would  naively expect. The physical reason is that the particle
spends more time close to $r_+$ where the approaching speed tends to zero.\\
In order to avoid possible misunderstanding, we remark that the model introduced in this section, is not equivalent to
classical BH solution of  Einstein equations. It has, however, something in common, i.e. the maximal extension of its
boundary is equal to the geometric Schwarschild radius.

\section{Quantization of the breathing horizon}
\label{qh}

Equation (\ref{etot}) is the starting point for the quantization of the system. As we were working in a relativistic
framework already at the classical level, the corresponding quantum equation will be of relativistic type as well.
By applying the canonical substitution rule

\begin{equation}
 p \longrightarrow i \frac{d}{dr}\ , \qquad (\hbar \equiv 1)
\end{equation}

we find the horizon wave equation 
 
\begin{equation}
  \frac{1}{r^2}\frac{d}{dr}\left(\, r^2 \frac{d\psi}{dr}\,\right) 
+ \left(\, E^2 - \frac{r^2}{4G_N^2}\,\right) \psi =0 \label{weq}
\end{equation}

where, $\psi$  is normalized as:

\begin{equation}
 4\pi \int_0^\infty dr r^2 \psi^\ast \psi =1 
\end{equation}

Since the intention is to interpret $\psi\left(\, r\,\right)$ as the wave-function  of a quantized Schwarzschild-like horizon, 
we have limited ourselves  to ``s-wave'' states only. It is in principle possible to allow  quantum
fluctuations with non-vanishing angular momentum, but since this is the first step towards a non-geometric quantum BHs, we limit
ourselves, in this paper, to the simplest possible case. We plan to treat the general  model in a future publication.
\\ 
At this point, several comments are in order. 
\begin{itemize}
\item One may wonder why equation (\ref{weq}) is not written in a Schwarzschild background geometry instead of flat space.
The answer is that we are not treating the back-reaction problem of a quantum particle  in a classical  Schwarzschild 
space-time. Rather, the ``particle'' \emph{is} the horizon itself, and $\psi(r)$ encodes the uncertainty in the horizon
radius.  
\item
A second reason justifying the form of the wave equation (\ref{weq}) is 
the holographic principle. The whole
quantum dynamics of the BH must be described in terms of a wave-function for the horizon only, and \emph{not} in terms 
of any bulk geometry, whatever it is.
\item
Finally, quantization naturally leads to a ''\emph{fuzzy}`` horizon which cannot be meaningfully described in terms of a classical
smooth surface.  The very distinction between the ''interior`` and ''exterior`` of the BH is no more  significant than
the distinction between the interior/exterior of a quantum wave-packet.
\end{itemize}

The solution turns out to be:

\begin{equation}
 \psi_n \left(\, r^2/2G_N\,\right) = N_n\, e^{-r^2/4G_N }\, L_n^{1/2}\left(\, r^2/2G_N\,\right)
\end{equation}

where

\begin{equation}
 N_n=\frac{\sqrt{n!}}{ 2\sqrt{\sqrt{2}\pi G_N^{3/2} \Gamma\left(\, n+3/2\,\right)}} \label{norm}
\end{equation}

The corresponding quantum BH mass spectrum \cite{Spallucci:2015jea,Spallucci:2014kua} is:

\begin{equation}
 E_n^2=\frac{1}{G_N}\left(\, 2n + \frac{3}{2}\,\right) \ ,\qquad n=0\ ,1\ ,2\ ,\dots
\label{spectrum}
\end{equation}

First thing to remark is the existence of a \emph{ground state} energy, or zero-point energy, near the Planck mass:

\begin{equation}
 E_{n=0}=1.22\times  M_{Pl.} \label{vacuum}
\end{equation}

Contrary to the semi-classical description where the mass can be arbitrary small, we find that in a genuine quantum description
the mass spectrum is bounded from below by $E_{n=0}$. In this model the quantization  solves the problem of
the ultimate stage of any process involving emission or absorption of energy.  Neither ''naked singularity`` nor empty Minkowski
space-time are allowed as final stage of the BH decay. The standard thermodynamical picture looses its meaning since
we are in a true quantum regime.\\
The excited states are equidistant much like in the case of an harmonic oscillator.\\
Having acquired the notion that Plankian BHs are quite different objects from their classical ``cousins'', we would like to address
the question of how to consistently connect Planckian and semi-classical BHs. As usual, one assumes that the quantum
system approaches the semi-classical one in the ``large-n'' limit in which the energy spectrum becomes continuous.
Before doing so, let us first consider the radial density describing the probability of finding the particle at distance $r$ 
from the origin, define as $ p(r)\equiv 4\pi r^2 \vert \psi \vert ^2 $:

\begin{eqnarray}
 && p_n(x)=\frac{2\, n!}{\Gamma\left(\, n +3/2\,\right)}\, x^2\, e^{-x^2}\, \left(\, L_n^{1/2}\left(\,x^2\,\right)\,\right)^2\ ,\\
 && x\equiv r/\sqrt{2G_N}
\end{eqnarray}

\begin{figure}[h!]
\includegraphics[height=5cm]{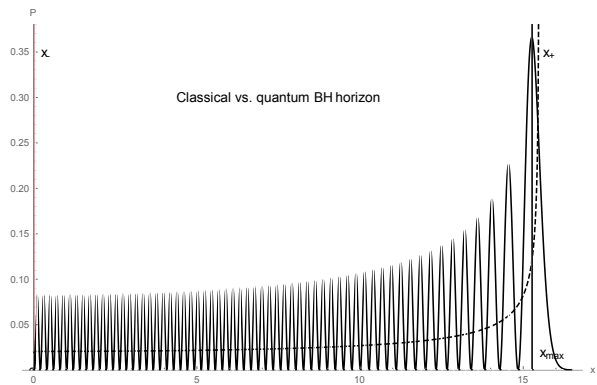}
\caption{Plot of the function $p_{n=60}(x)$,  (continuous line) vs classical probability (dashed line). For large $n$ the
position of the  peak approaches the classical Schwarzschild radius $r_+= 2E G_N$.}
\label{massimi}
\end{figure}

The local maxima in figure(\ref{massimi}) represent the most probable size of the Planckian BH.  
These maxima  are solutions of the equation

\begin{equation}
\left(\, 1 -x^2 +4n\,\right)\, L_n^{1/2}\left(\, x^2\,\right)-2\left(\, 2n +1/2\,\right)\, 
L_{n-1}^{1/2}\left(\, x^2\,\right)=0 \label{maxima}
\end{equation}

Equation (\ref{maxima}) cannot be solved analytically , but its large-$n$ limit can be evaluated as follows.
First, perform the division $L_{n}^{1/2}/L_{n-1}^{1/2}$, and then write

\begin{equation}
 L_n^{1/2}\left(\,x^2\,\right)=
P_2\left(\, x^2\,\right)\, L_{n-1}^{1/2}\left(\, x^2\,\right)+R_{n-2}\left(\, x^2\,\right)\label{ratio}
\end{equation}

where, 

\begin{eqnarray}
&& P_2 =\frac{a_n}{b_{n-1}}\left(\, x^2-2n+1/2\, \right)\\
&& R_{n-2}=c_{n-2} x^{2n-4}+\dots\ , \nonumber\\
&&\>\>\>\>\> \>\>\>\>\>\>=-\left(\, n-1\,\right)\left(\, n-1/2\,\right)\, a_n\,x^{2n-4}+\cdots
\end{eqnarray}

By inserting equation (\ref{ratio}) in equation (\ref{maxima}) and by keeping terms up order $x^{2n-2}$, the equation
for maxima turns into

\begin{eqnarray}
&& \left[\,x^2-2n-1\,\right]\,\left[\, x^2-2n+1/2\,\right]+\left[\, 2n+1\,\right]\frac{b_{n-1}}{a_n}=\nonumber\\
 &&  =(n-1)(n-1/2)
\label{max1}
\end{eqnarray}

where the coefficients  are given by

\begin{eqnarray}
&& a_n=\frac{(-1)^n}{n!}\ ,\\
&& b_{n-1}=\frac{(-1)^{n-1}}{(n-1)!}
\end{eqnarray}

Equation (\ref{max1}), for large $n$ reduces to
 \begin{eqnarray}
&& 3n^2= \left(\,x^2-2n\,\right)^2 \nonumber\\
&& x^2= 2n+\sqrt{3} n=3.73n\nonumber
\end{eqnarray}

On the other hand, the classical radius of the horizon is obtained as

\begin{equation}
  \frac{r_+^2}{2G_N}= 2G_NE^2  = 4n 
 \end{equation}
 
which leads to

\begin{equation}
 x^2_+=4n
\end{equation}
 
Thus, we find that most probable value of $r$ approaches the horizon radius $r_+$ for $E>> M_p$, 
restoring the (semi)classical picture of BH.

\section{Conclusion}
\label{end}

In this paper we have presented the dynamics of a genuine quantum BH in a non-geometrical framework.  First, in the absence of
non-geometrical description of classical BHs, we have envisaged a classical model for an  harmonically oscillating 
spherical surface whose maximum radius turns out to be the classical Schwarzschild radius.  This model was necessary as a
starting point to obtain its quantum version through a canonical quantization procedure. We would like to stress that the world
``classical'' refers still to a small object compared to cosmic BHs of astronomical size. In this way we have realized a dynamical
that naturally incorporates the qualitative ideas underlying the non-geometrical approach by Dvali and co-workers.
To be more precise one can identify their characteristic occupational number $N$ with our principal quantum number $n$ in
(\ref{vacuum}) as 
\begin{equation}
 N=4n
\end{equation}
This identification allows to set the analogy between graviton BEC and excited states in our model. Equation  (\ref{vacuum})
describes a spectrum of equidistant energy levels as in the quantum harmonic oscillator case. This admits a re-interpretation
as \emph{many-body} system, e.g. ``$N$-gravitons'' rather than a single-particle spectrum.\\
Furthermore, both in our case as well as in \cite{Dvali:2011aa} the mass of the QBH is 

\begin{equation}
 E \propto \sqrt{N}
\end{equation}

and the quantum behavior is characterized by a single integer $N$, or $n$, counting the number of ``constituents''.\\

Our model confirms, in a surprisingly simple manner, the growing believe
that Planckian scale BHs behave in a  profoundly different way from classical gravitationally collapsed objects.
Finally, an unexpected feature of the quantum behavior is to turn the ''dreadful``  classical BHs into an harmless  
quantum  ``black'' particles.

\end{document}